\documentclass{emulateapj}
\usepackage{graphicx}
\usepackage[usenames,dvips]{color}

\newcommand{\be}{\begin{equation}}
\newcommand{\bel}[1]{\begin{equation}\label{eq:#1}}
\newcommand{\ee}{\end{equation}}
\newcommand{\bd}{\begin{displaymath}} 
\newcommand{\ed}{\end{displaymath}}   
\newcommand{\bea}{\begin{eqnarray}}
\newcommand{\beal}[1]{\begin{eqnarray}\label{eq:#1}}
\newcommand{\eea}{\end{eqnarray}}

\newcommand{\eqref}[1]{\ref{eq:#1}}

\newcommand{\lsim }{{\lower0.8ex\hbox{$\buildrel <\over\sim$}}}
\newcommand{\gsim }{{\lower0.8ex\hbox{$\buildrel >\over\sim$}}}

\def\apj{ ApJ}
\def\aap{ A\&A}

\def\araa{ARAA}
\def\aj{AJ}
\def\apjs{ApJ Supp}

\def\Chandra{${\it Chandra}$}

\def\XMM{{\it XMM-Newton}}
\def\ergss{ergs s$^{-1}$}

\def\simge{\mathrel{%
   \rlap{\raise 0.511ex \hbox{$>$}}{\lower 0.511ex \hbox{$\sim$}}}}
\def\simle{\mathrel{
   \rlap{\raise 0.511ex \hbox{$<$}}{\lower 0.511ex \hbox{$\sim$}}}}

\newcommand{\Msun}{\ifmmode {M_{\odot}}\else${M_{\odot}}$\fi}
\newcommand{\Lsun}{\ifmmode {L_{\odot}}\else${L_{\odot}}$\fi}
\newcommand{\Rsun}{\ifmmode {R_{\odot}}\else${R_{\odot}}$\fi}

\shorttitle{Cooling of Cas A Neutron Star}
\shortauthors{Heinke \& Ho}

\begin{document}
\title{Direct Observation of the Cooling of the Cassiopeia A Neutron Star}  

\author{Craig~O. Heinke\altaffilmark{1}, Wynn C. G. Ho\altaffilmark{2} }

\altaffiltext{1}{Dept. of Physics, University of Alberta, Room 238 CEB, Edmonton, AB T6G 2G7, Canada; heinke@ualberta.ca}
\altaffiltext{2}{School of Mathematics, University of Southampton, Southampton SO17 1BJ, UK; wynnho@slac.stanford.edu}


\begin{abstract}
The cooling rate of young neutron stars gives direct insight into their internal makeup.  
Although the temperatures of several young neutron stars have been measured, until now a young 
neutron star has never been observed to decrease in temperature over time.  
We fit 9 years of archival \Chandra\ ACIS spectra of the likely neutron star in the $\sim$330 years old Cassiopeia~A supernova remnant with our non-magnetic carbon atmosphere model.  
Our fits show a relative decline in the surface temperature by 4\% (5.4$\sigma$, from $2.12\pm0.01\times10^6$~K in 2000 to $2.04\pm0.01\times10^6$~K in 2009) and observed flux (by 21\%).
Using a simple model for neutron star cooling, we show that this temperature decline could indicate that the neutron star became isothermal sometime between 1965 and 1980, and constrains some combinations of neutrino emission mechanisms and envelope compositions.
However, the neutron star is likely to have become isothermal soon after formation, in which case the temperature history suggests episodes of additional heating or more rapid cooling.
Observations over the next few years will allow us to test possible explanations for the temperature evolution.
\end{abstract}


\maketitle

\section{Introduction}\label{s:intro}

The internal composition and structure of neutron stars (NSs) remains unclear \citep[e.g.][]{Lattimer04}.  Areas of uncertainty include whether exotic condensates occur in the NS core, the symmetry energy and thus proton fraction in the core, the behavior of superfluidity among neutrons and protons, the conductivity of the NS crust, and the chemical composition of the outer envelope. NSs are heated to billions of degrees during supernovae, and cool via a combination of neutrino and photon emission.   Observing the cooling rates of young NSs is a critical method to constrain the uncertainties \citep[see][for reviews]{Tsuruta98,Yakovlev04,Page06}.

To date, observations of young cooling NSs have been restricted to measuring the temperature of individual NSs at one point in time.  As NSs may differ in their mass, envelope composition, etc., a measurement of the cooling rate of a young NS is needed to determine its cooling trajectory.  Since neutrino radiation (rather than the observed photon radiation) is the dominant source of cooling during the first $\sim10^5$ years, measurements of cooling rates during this time require measuring a temperature decline over time.  No young NS has previously been observed to cool steadily over time. Though the $\sim 10^6$-years-old NS RX~J0720.4$-$3125 has shown temperature variations of $\sim$10\% over $\approx 7$ years \citep{deVries04,Hohle09}, this variation is ascribed to either a glitch-like event or precession of surface hot spots \citep{Haberl06,vanKerkwijk07,Hohle09}.  Magnetars, such as 4U~0142$+$61, have shown temperature variations, along with changes in their pulsed fraction and pulse profile \citep{Dib07}, but these are likely due to magnetic field reconfiguration events.  

The compact central object at the center of the Cassiopeia A (Cas~A) supernova remnant was discovered in \Chandra's first-light observations \citep{Tananbaum99}, and quickly identified as a likely NS, which we assume here.  It is presently the youngest-known NS, as the remnant's estimated age is $\approx 330$ years \citep{Fesen06}.  It is relatively close-by \citep[$d=3.4^{+0.3}_{-0.1}$ kpc,][]{Reed95} and the supernova remnant has been well-studied, with over a megasecond of \Chandra\ ACIS observations spread over 10 years \citep{Hwang04,Delaney04,Patnaude07,Patnaude09}.  However, its spectrum (modeled as a blackbody or a magnetic or non-magnetic hydrogen atmosphere) was inconsistent with emission from the full surface of the NS \citep{Pavlov00,Chakrabarty01,Pavlov09}.  Timing investigations using the \Chandra\ HRC and \XMM\ have failed to identify pulsations down to a pulsed fraction level of $<$12\% \citep{Murray02,Mereghetti02,Ransom02,Halpern10}, indicating that the emission is probably from the entire surface.  These apparently contradictory observations are reconciled by the discovery that an unmagnetized ($B<10^{11}$ G) carbon atmosphere provides a good fit to the \Chandra\ ACIS data, with the emission arising from the entire surface of the Cas A NS \citep{Ho09}.  

\citet{Pavlov04} examined two long ACIS observations (50 ks each) of the Cas~A NS from 2000 and 2002, along with several short (2.5 ks) calibration observations, finding no significant changes in flux.  Upon re-examination of archival {\it Einstein} and {\it Rosat} data, the NS was only barely detected, and thus could not be used to search for variability \citep{Pavlov00}. \citet{Pavlov09} mention that the flux measured in their 2006 observation is slightly lower than reported previously, but do not attempt to determine whether the difference is real.  Before \citet{Ho09}, it was not expected that the emission arises from the entire surface of the NS, so further serious searches for temperature variations were not undertaken.  Here we utilize the full \Chandra\ ACIS archive of Cas~A NS observations to measure the temperature changes from 2000 to 2009.

\section{X-ray Analysis}\label{s:X-ray}

We analyzed all \Chandra\ ACIS-S exposures without gratings, longer than 5 ks, of Cas A, listed in Table 1.  We also analyzed the zeroth-order grating spectrum from ObsID 1046, which was taken in 2001.  Although the fit to a carbon atmosphere model spectrum was good, the derived temperature ($\log T_\mathrm{s}=6.282^{+.004}_{-.004}$) 
is significantly lower than all other ACIS measurements. This is likely due to calibration differences between the zeroth-order grating observation and observations without gratings.  Similar cross-calibration uncertainties prohibit direct comparison of \Chandra\ HRC or \XMM\ observations with \Chandra\ ACIS observations.  The HRC-I observations lack spectral information; the HRC-I team also uses the Cas~A NS as a quantum efficiency calibration source\footnote{cxc.harvard.edu/ccr/proceedings/07\_proc/presentations/possonbrown3/}, which we suspect may be negatively impacting the HRC-I calibration.  \XMM\ observations suffer substantially increased background from the supernova remnant.

ObsID 6690 was taken using a subarray mode to alleviate the effects of pileup on the Cas~A NS; all other data were taken in full-frame mode, with frame times of 3.04 or 3.24 s.  We used CIAO 4.2 (with CALDB 4.2.1) to reprocess the observations with current calibrations, extract spectra and create responses.  We used a 4-pixel (2.37'') radius region for source extraction,  and an annulus from 5 to 8 pixels for background.  Our source region is slightly larger than used by \citet{Pavlov09} and \citet{Ho09}, giving a more complete flux estimate when the point-spread function is asymmetric (the NS was slightly off-axis in most observations).  Most data were taken in GRADED mode, so (apart from ObsID 6690) we could not correct the data for charge-transfer inefficiency.  The time-dependent ACIS quantum efficiency degradation is modeled, but has a small effect on our analysis due to the few counts below 1 keV.

The nature of our analysis requires that we consider possible instrumental effects on the effective area carefully.  We have identified the following possible effects: contaminant effects on low-energy quantum efficiency (QE); increasing charge-transfer inefficiency effects on QE; bad pixels/columns; and pileup.

Charge-transfer inefficiency can alter the ``grade'' designation of events from ``good'' (likely X-rays) to ``bad'' (usually cosmic rays) grades, leading to deletion of good events.  However, the Chandra X-ray Center maintains accurate calibration files for the S3 chip without CTI correction (acisD2000-01-29qeuN0005.fits), which addresses GRADED mode data, and in any case the QE is significantly affected only below 1 keV\footnote{CXC HelpDesk ticket 12871.}.  A molecular contaminant has been building on the ACIS detector, reducing the QE at 1 keV by $\sim$10\% over the \Chandra\ mission\footnote{http://cxc.harvard.edu/cal/Acis/Cal\_prods/qeDeg}.  However, the effects of this contaminant are negligible above 2 keV, and are now calibrated across the detector and through time; uncertainties in changes of effective area with time are believed to be $<$3\% for energies above 0.7 keV.  If this affected our data, we should expect to see greater variations in the lower-energy than higher-energy data, which is not the case (see below).

Bad pixels may affect the inferred QE by removing good 
data\footnote{http://cxc.harvard.edu/cal/Acis/Cal\_prods/badpix/index.html}. 
A bad pixel region at CHIPX=495-499, and the node boundary at CHIPX=512-513, were crossed by the dither pattern of the Cas A NS in the Hwang datasets in April/May 2004.  The CHIPX=496-498 bad pixels were not telemetered to the ground (bias values of 4095), so it is not possible to check the effective area calibration by extracting data including the bad pixels.  We analyzed this data using the appropriate bad pixel lists, and though the average temperatures were within the range of our other data, we saw significant ($\sim2$\%) variations in the fitted temperatures correlated with changes in the sky position of the bad pixels on short ($\sim$week)  timescales.  We suspect these changes are due to the responses incompletely adjusting for the effects of the bad pixels, and therefore do not include data with bad pixel regions crossing the Cas A NS in our study of the temperature variations (though we list the results for completeness).

Pileup is the recording of two photons during one frametime as one event, leading to changes in the spectrum and the rejection of some events due to a change in their grade \citep{Davis01}.  Although the absolute effects of pileup remain uncertain, most of our observations suffer the same level of pileup.  ObsID 6690 suffers much less pileup due to its short (0.3 s) frame time; the first 3 observations have a slightly longer frame time (3.24 s vs. 3.04 s).  Thus we suspect that ObsID 6690 may have a systematic shift in T compared to the other observations (though it is probably the most accurate in an absolute sense); we show results from ObsID 6690 but exclude it from fits to the temperature trends.

After some experimenting, we chose to group 50-ks datasets by 200 counts, and to increase the grouping for longer exposures to produce a similar number of bins.  We merge ObsID 9117 with 9773, and ObsID 10935 with 12020, as the datasets are short ($\sim$25 ks) and adjacent in time.  
We use a similar spectral fit to that in \citet{Ho09}, a model containing photoelectric absorption (with abundances from \citealt{Wilms00}), an unmagnetized carbon NS atmosphere, and scattering by interstellar dust \citep{Predehl03}, all convolved with the Davis pileup model \citep{Davis01} in XSPEC.  We fix the distance to 3.4 kpc. 

There is degeneracy among the fit parameters for the NS mass $M$ and radius $R$, interstellar absorption $N_\mathrm{H}$, and surface temperature $T_\mathrm{s}$.  Since $M$,  $R$ and $N_\mathrm{H}$ are not expected to vary between observations, these are held constant at the best-fit values in order to explore variations in $T_\mathrm{s}$.  If we exclude the bad-pixel-affected data, the best-fit gives $M=2.01$ \Msun, $R=8.3$~km, and $N_\mathrm{H}=1.82\times10^{21}$, while the best-fit including this data gives $M=1.65$ \Msun, $R=10.3$~km, and $N_\mathrm{H}=1.74\times10^{21}$.
We use the best-fit values for $M$, $R$ and $N_\mathrm{H}$ derived from including all the data, as this provides substantially more information on the spectral curvature (and a more believable NS mass), but we do not consider the temperatures derived from the questionable data in our analysis of temperature variations.  Using the best-fit values without the bad-pixel-affected data gives similar results to those described below, except that the temperatures are all shifted slightly higher by about the same amount.  The allowed range of absolute temperatures, and the relationships between the fit parameters are explored in detail in Yakovlev et~al.\ (in prep).
We allow the grade migration parameter $\alpha$ in the \citet{Davis01} pileup model to vary between observations with different frame times, giving values of  $\alpha=0.27\pm0.06$ for the 3.24 s frame time observations, $\alpha= 0.24\pm0.05$ for the 3.04 s observations, and $\alpha<0.62$ for ObsID 6690 (all 90\% confidence).  Our results are not substantially changed by requiring $\alpha$ to be fixed across all observations.  We quote $1\sigma$ confidence errors for $T_\mathrm{s}$ and $L_\mathrm{bol}$ in Table 1, for convenience in fitting the temperature variations.  The observed flux change ($\sim$21\% in absorbed flux) does not depend on the choice of spectral model.  

We find that requiring the same NS temperature for the (non-bad-pixel-affected) observations produces a poor $\chi^2$ (= 184 for 131 degrees of freedom) with a null hypothesis probability of $1.6\times10^{-3}$.  Allowing the NS temperature to vary reduces the $\chi^2$ to 113 for 126 degrees of freedom, which gives a probability of 79\% for an acceptable fit.  An F-test finds a statistic of 15.8 and probability of $4\times10^{-12}$, indicating that the additional parameters substantially improve the fit.  Figure~1 shows the best-fit spectral fit with a constant temperature (excluding the bad-pixel-affected data).  Substantial differences in the data/model ratio are clearly evident; these differences increase slightly at higher energies (as expected for temperature variations) from early to later observations.  If the molecular contaminant or charge-transfer inefficiency were the primary cause of the variation, we would expect the data/model ratio to vary principally at low energies, which is not seen.

\begin{figure}
\figurenum{1}
\includegraphics[angle=-90,scale=.35]{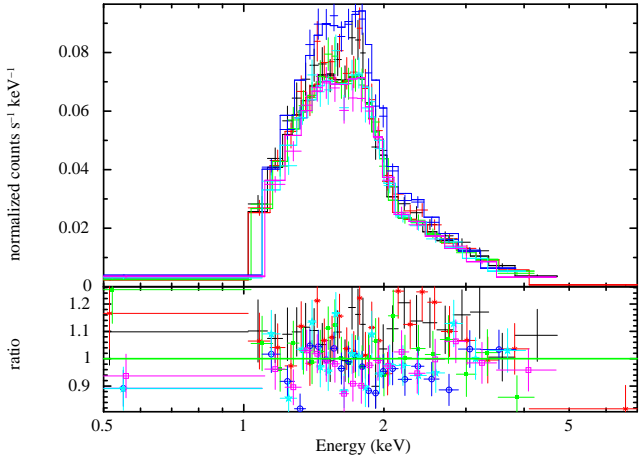}
\caption[June11_dat+ratio.eps]{ \label{fig:ratio}
Illustrative spectral fit of \Chandra\ ACIS data to our non-magnetic carbon atmosphere model spectrum, with all temperatures forced to be equal.  Upper panel shows the data and model.  Lower panel shows the ratio of data/model, with the different datasets marked: 2000: (black) plain crosses, 2002: (red) asterisks, Feb. 2004: (green) filled squares, 2006: (blue) circles, merged 2007: (cyan) stars, merged 2009: (magenta) open squares.  The 2006 data has a higher countrate (upper panel) due to its lower pileup fraction.  The change of countrate from early (2000-2002) to later (2006-2009) spectra, and its spectral uniformity, can be clearly seen in the ratio plot. 
} 
\end{figure}

\begin{figure}
\figurenum{2}
\includegraphics[angle=0,scale=.43]{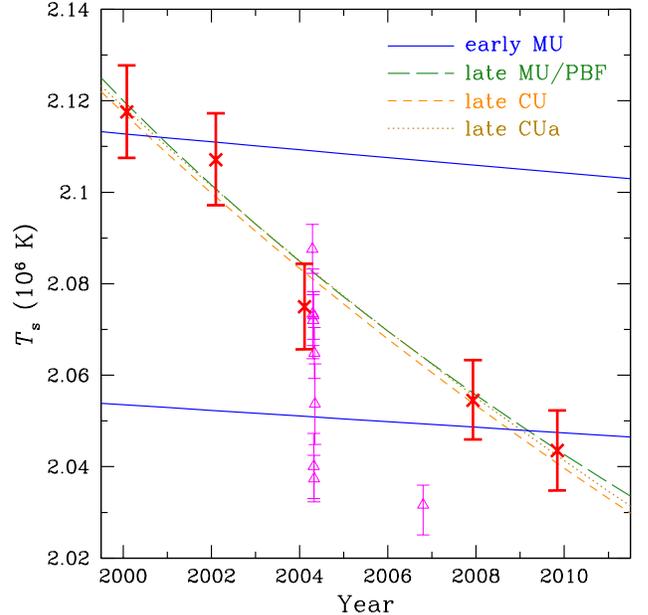}
\caption[fitTvst.eps]{ \label{fig:decline}
Surface temperature $T_\mathrm{s}$ of the Cas A NS, obtained from spectral fits of \Chandra\ observations, as a function of time (crosses and triangles indicate best-fit values with $1\sigma$ errorbars).  Curves are fits (to well-calibrated data, the 5 red crosses; see text) with a simple model for NS cooling by modified Urca or pair breaking and formation (MU/PBF; long-dashed) or condensate Urca with an iron envelope (CU; short-dashed) or a fully accreted light-element envelope (CUa; dotted) after a long delay in thermal relaxation, while the solid curves are for cooling by modified Urca with relaxation shortly [$\approx 100$~y (upper) and $\approx 20$~y (lower)] after NS formation (see text).
} 
\end{figure}

Figure~\ref{fig:decline} shows the temperature variation over the almost ten years the Cas~A NS has been observed by \Chandra.  We show the five well-calibrated temperature measurements (as well as, for reference, the 2006 subarray measurement and the 2004 measurements affected by bad pixels).
A clear decrease of $3.6\pm0.6\%$ in $T_\mathrm{s}$ ($15\pm4\%$ decrease in bolometric luminosity) is evident during this period.  
temperature drop of $1.5\pm0.5$\% within only 21 days.  
This gives a temperature evolution timescale $T_\mathrm{s}/(\Delta T_\mathrm{s}/\Delta t)$ of $\sim 280$~y.

\section{Discussion}\label{s:discuss}

A detailed analysis of the Cas~A NS temperature evolution is beyond the scope of this paper.  Work is underway on investigating the thermal history of the Cas~A NS in the context of neutrino cooling calculations (Yakovlev et al., in prep.).
Here we briefly describe NS cooling theory and provide a simplified model
to compare with the observed evolution
\citep[see][for review]{Tsuruta98,Yakovlev04,Page06}.

The long-term thermal history of a NS is determined by the neutrino luminosity and heat capacity of the core and the composition (i.e., thermal conductivity) of the surface layers.
At very early times, the core cools rapidly via neutrino emission while the temperature of the thermally-decoupled crust remains nearly constant.  A cooling wave travels from the core to the surface, bringing the NS to a relaxed, isothermal state.  Depending on the properties of the crust, the relaxation time can take
$\sim 10-100$ y \citep{Lattimer94,Gnedin01}.
For the next $\sim 10^5-10^6$ y, surface temperature changes reflect changes in the interior temperature as neutrino emission continuously removes heat from the star.

Let us assume that the Cas~A NS has become thermally-relaxed and that the
observed temperature decline is due solely to neutrino emission.  The thermal
evolution of a young NS is then governed by the thermal balance equation
$C(dT/dt)\approx -L_{\nu}$, where $T$ is the interior temperature, $C$ is the
total heat capacity, and $L_{\nu}$ is the total neutrino luminosity.
The ratio $L_{\nu}/C$ thus determines the rate of temperature change.
The heat capacity for a non-superfluid NS is
$C\sim 10^{38}\,T_8\mbox{ ergs K$^{-1}$}$, where $T_8=T/10^8$~K.
For simplicity, we consider $L_{\nu}$ to be given by a single neutrino process,
i.e., either the slow modified Urca (MU) process with
$L_{\nu}^\mathrm{MU}\sim 10^{32}\,T_8^8\mbox{ ergs s$^{-1}$}$
or a fast condensate Urca (CU) process with
$L_{\nu}^\mathrm{CU}\sim 10^{36}\,T_8^6\mbox{ ergs s$^{-1}$}$
\citep[see, e.g.,][for review]{Yakovlev04,Page06};
note that the nucleon direct Urca process would result in a temperature
below that observed within one year after becoming isothermal.
The evolution equation then results in $T_8(t)\approx T_0(\eta t)^{-n}$, where
$t$ is in years, $\eta$ encapsulates the uncertainties in the coefficient
of $L_{\nu}/C$, and
$n^\mathrm{MU}=1/6$ and $T_0^\mathrm{MU}=9$
and $n^\mathrm{CU}=1/4$ and $T_0^\mathrm{CU}=3$.
Note that we assume the current temperature is much lower than the initial
temperature.
To convert the evolution of the interior temperature to one for the surface
temperature, we use the relation
$T_{\mathrm{s}6}\approx 1.1\,T_8^{11/20}$ for an iron envelope
\citep{Gudmundsson82} and
$T_{\mathrm{s}6}\approx 1.8\,T_8^{17/28}$ for a (fully accreted)
light-element envelope \citep[][see also \citealt{Potekhin03}]{Potekhin97},
where $T_{\mathrm{s}6}=T_\mathrm{s}/10^6$~K.
We thus obtain
\be
T_{\mathrm{s}6}(t)= T_{\mathrm{s}0}(\eta t)^{-\alpha},
\label{eq:ts}
\ee
where $\alpha^\mathrm{MU}=11/120$ and $T_{\mathrm{s}0}^\mathrm{MU}=4$
and $\alpha^\mathrm{CU}=11/80$ and $T_{\mathrm{s}0}^\mathrm{CU}=2$
for an iron envelope and
$\alpha^\mathrm{MUa}=17/168$ and $T_{\mathrm{s}0}^\mathrm{MUa}=7$
and $\alpha^\mathrm{CUa}=17/112$ and $T_{\mathrm{s}0}^\mathrm{CUa}=4$
for a fully accreted envelope.

In Fig.~\ref{fig:decline}, we show the predictions of the temperature decline given by
eq.~(\ref{eq:ts}), where $t\equiv(\tau-\tau_\mathrm{x})$, $\tau$ is
the year, and $\tau_\mathrm{x}$ is the approximate year when the star
becomes isothermal (after which our simple thermal evolution scaling is valid).
The deviation from unity of $\eta$ (which encompasses the uncertainties
in our knowledge of $L_{\nu}/C$) is a rough measure of the likelihood that
a given process and composition is responsible for the temperature evolution
seen in the Cas~A NS.
If we assume the NS took a long time to thermally relax (so that it only
became isothermal recently), a fit to the observations (excluding the
bad-pixel-affected data and the 2006 subarray mode data) for each neutrino
emission process and envelope composition yields $\chi^2\approx 1.4$
for 3 degrees of freedom.
For slow cooling and an accreted envelope, we find $\eta$(MUa)$=6000$,
which requires a significantly higher neutrino luminosity
and/or lower heat capacity than traditionally considered;
in other words, the observed $T_\mathrm{s}$ requires a $T$ that is too low
for slow processes to achieve in $300$~y.
On the other hand, $\eta$(MU)$=50$, $\eta$(CU)$=0.02$, and $\eta$(CUa)$=2$,
and thus are possibile scenarios.
We have not taken into account the large (suppression) effects of superfluidity
on the neutrino luminosity and nucleon heat capacity because of the
relatively-unknown critical temperature at which these effects begin to occur
\citep[see][and references therein]{Page04}.
However, we note that if strong superfluid pairing exists in the NS,
then neutrino emission by pair breaking and formation (PBF) produces
$L_{\nu}^\mathrm{PBF}\gtrsim 10L_{\nu}^\mathrm{MU}$ \citep{Gusakov04};
this results in $\eta$(PBF)$\lesssim 5$.
We also find $\tau_\mathrm{x}$(MU/PBF)=1980, $\tau_\mathrm{x}$(CU)=1968,
and $\tau_\mathrm{x}$(CUa)=1964,
i.e., $\sim 300$~y after the supernova.  Despite the good fit to the
data, the very long relaxation time makes this scenario questionable.

We also show two (MU) cooling curves in Fig.~\ref{fig:decline}, which
assume that the NS became thermally-relaxed early ($\approx 20$~y and
$\approx 100$~y) after formation; no fit is done, and $\eta\approx 5$
is assumed.
In this case,
the clear deviation from a single neutrino cooling curve suggests that
a transient heating (or cooling) episode occurred.
Transient heating can originate from external causes, e.g., accretion of circumstellar gas or asteroids \citep[e.g.][]{Jura03,Cordes08}, or from internal sources, e.g., by the same mechanism that is responsible for pulsar glitches \citep[][and references therein]{vanRiper91,Seward00,Helfand01}.  Further monitoring of the Cas A NS temperature evolution should allow us to test the possibility of transient heating via a return to quiescent neutrino cooling.

\acknowledgements

We are grateful to D. Patnaude for allowing us to use the 2009 \Chandra\ observations during his proprietary period, and to D.I. Jones for asking the question that motivated this work.  
We thank G. G. Pavlov for pointing out the issue with bad pixels affecting the Hwang data.
We thank D. Yakovlev for discussions, and N. Bonaventura at the CXC Help Desk for addressing calibration questions.  
We acknowledge the use of public data from the \Chandra\ data archive.  
WCGH appreciates the use of the computer facilities at the Kavli
Institute for Particle Astrophysics and Cosmology.   COH acknowledges the support of NSERC, and WCGH acknowledges support from the Science and Technology Facilities Council (STFC) in the United Kingdom through grant number PP/E001025/1.

{\it Facilities:} \facility{CXO (ACIS)}

\bibliographystyle{apj}


\begin{deluxetable}{lcccccc}
\tablewidth{7.0truein}
\tablecaption{\textbf{Spectral Fits to Cas A Neutron Star}}
\tablehead{
{\textbf{Start time}} & {\textbf{MJD}} & {\textbf{Exposure}} & {\textbf{Frame time}} & {$\log T_\mathrm{s}$} & {$L_\mathrm{bol}$, 1 eV-10 keV} & {\textbf{ObsID}} \\
 (UT) &   & ks & s & K  & ($\times10^{33}$ \ergss) &   }
\startdata
\multicolumn{7}{c}{Directly comparable data} \\
2000-01-30 10:40 & 51573.74 & 50.56 & 3.24 & 6.3258$^{+0.0021}_{-0.0021}$ & $7.95^{+0.15}_{-0.15}$ & 114 \\
2002-02-06 06:22 & 52311.56 & 50.3   & 3.24 & 6.3237$^{+0.0021}_{-0.0021}$ & $7.79^{+0.15}_{-0.15}$ & 1952 \\
2004-02-08 17:41 & 53044.03 & 50.16 & 3.24 & 6.3170$^{+0.0020}_{-0.0020}$ & $7.33^{+0.13}_{-0.13}$ & 5196 \\
2007-12-05 22:00 & 54441.36 & 25.18 & 3.04 & 6.3127$^{+0.0018}_{-0.0018}$ & $7.05^{+0.12}_{-0.11}$ & 9117 \\
2007-12-08 12:34 & $^a$ & 25.17 & 3.04 & $^a$ & $^a$ & 9773 \\
2009-11-02 22:24 & 55138.59 & 23.58 & 3.04 & 6.3104$^{+0.0019}_{-0.0019}$ & $6.90^{+0.12}_{-0.11}$ & 10935 \\
2009-11-03 22:32 & $^b$ & 22.68 & 3.04 & $^b$ & $^b$ & 12020 \\
\hline
\multicolumn{7}{c}{Other data}\\
\hline
2004-04-14 19:47 & 53110.79 & 166.72 & 3.04 & 6.3196$^{+0.0011}_{-0.0011}$ & $7.50^{+0.07}_{-0.05}$ & 4638 \\
2004-04-18 21:18 & 53114.14 & 42.84  & 3.04 & 6.3167$^{+0.0020}_{-0.0021}$ & $7.30^{+0.12}_{-0.12}$ & 5319 \\
2004-04-20 08:41 & 53116.20 & 145.38 & 3.04 & 6.3164$^{+0.0012}_{-0.0012}$ & $7.28^{+0.07}_{-0.07}$ & 4636 \\
2004-04-22 18:40 & 53118.74 & 165.66 & 3.04 & 6.3166$^{+0.0011}_{-0.0011}$ & $7.30^{+0.07}_{-0.07}$ & 4637 \\
2004-04-25 09:37 & 53120.87 & 80.13   & 3.04 & 6.3097$^{+0.0015}_{-0.0015}$ & $6.85^{+0.09}_{-0.09}$ & 4639 \\
2004-04-28 05:43 & 53124.11 & 150.59 & 3.04 & 6.3091$^{+0.0011}_{-0.0011}$ & $6.82^{+0.06}_{-0.06}$ & 4634 \\
2004-05-01 00:44 & 53126.82 & 136.82 & 3.04 & 6.3149$^{+0.0012}_{-0.0012}$ & $7.19^{+0.07}_{-0.07}$ & 4635 \\
2004-05-05 22:59 & 53131.28 & 55.11   & 3.04 & 6.3125$^{+0.0018}_{-0.0019}$ & $7.03^{+0.11}_{-0.11}$ & 5320 \\
2006-10-19 08:18 & 54027.75 & 70.18 & 0.34 & 6.3079$^{+0.0009}_{-0.0014}$ & $6.74^{+0.06}_{-0.08}$ & 6690 \\
\enddata
\tablecomments{Carbon atmosphere spectral fits, with fixed distance of 3.4 kpc, neutron star mass of 1.648~\Msun, radius of 10.3~km, and $N_\mathrm{H}=1.74\times10^{21}$ cm$^{-2}$ (best-fit).  
Temperature and luminosity errors are $1\sigma$ confidence for a single parameter.  MJD values are for the midpoints of the observations, or the weighted midpoints of merged sets.  \\ 
$^a$ Combined with ObsID 9117; values for MJD, $\log T$, and $L_\mathrm{bol}$ are for the merged set.
$^b$ Combined with ObsID 10935; values for MJD, $\log T$, and $L_\mathrm{bol}$ are for the merged set. 
}
\end{deluxetable}


\end{document}